\documentclass[letter]{aa} 

\usepackage{graphicx}
\usepackage{txfonts}
\usepackage[colorlinks]{hyperref}

\def\aap{A\&A}
\def\apj{ApJ}
\def\apjs{ApJS}

\def \hi {\ion{H}{i}}
\def\h2{H$_2$}

\def\kms{km\,s$^{-1}$}

\def\deg{\hbox{$^\circ$}}
\def\arcmin{\hbox{$^\prime$}}
\def\arcsec{\hbox{$^{\prime\prime}$}}

\def\farcm{\hbox{$.\mkern-4mu^\prime$}}

\defcitealias{Kalberla2021}{Paper~I}
\defcitealias{Kalberla2023}{Paper~II}
\defcitealias{Kalberla2024}{Paper~III}

\begin{document}

\title{The cold neutral medium in filaments at high Galactic latitudes}


\subtitle{}

   \author{P.\ M.\ W.\ Kalberla }

\institute{Argelander-Institut f\"ur Astronomie, University of Bonn,
           Auf dem H\"ugel 71, 53121 Bonn, Germany \\
           \email{pkalberla@astro.uni-bonn.de}
 }

   \authorrunning{P.\,M.\,W. Kalberla } 

   \titlerunning{Cold neutral medium in filaments }

   \date{Received 27 October 2024 / Accepted  29 January 2025 }

  \abstract 
{The \hi\ distribution at high Galactic latitudes has been found to be
  filamentary and closely related to the far infrared (FIR) in caustics
  with coherent velocity structures. These structures trace the
  orientation of magnetic field lines.}
{Recent absorption observations with the Australian SKA Pathfinder
  Telescope have led to major improvements in the understanding of the
  physical properties of the cold neutral medium (CNM) at high Galactic
  latitudes. We use these results to explore how far the physical state
  of the CNM may be related with caustics in \hi\ and FIR.  }
{We traced filamentary FIR and \hi\ structures and probed the absorption data
  for coincidences in position and velocity. }
{ Of the absorption positions, 57\% are associated with known
  FIR/\hi\ caustics, filamentary dusty structures with a coherent
  velocity field.  The remaining part of the absorption sample is
  coincident in position and velocity with genuine \hi\ filaments that
  are closely related to the FIR counterparts. Thus, within the current
  sensitivity limitations, all the positions with
  \hi\ absorption lines are associated with filamentary structures in
  FIR and/or \hi. We summarize the physical parameters for the CNM along
  filaments in the framework of filament velocities $v_\mathrm{fil}$
  that have been determined from a Hessian analysis of FIR and
  \hi\ emission data. Velocity deviations between absorption components
  and filament velocities are due to local turbulence, and we determine
  for the observed CNM an average turbulent velocity  dispersion of
    $ 2.48 < \delta v_\mathrm{turb} < 3.9 $ \kms.  The CNM has a mean
   turbulent Mach number of $M_\mathrm{t} = 3.4 \pm 1.6 $ \kms. }
{Most, if not all, of the CNM in the diffuse interstellar medium at high Galactic
  latitudes is located in filaments, identified as caustics with the
  Hessian operator.  }


  \maketitle
%

\section{Introduction}
\label{Intro}

High-resolution large-scale surveys in \hi\ emission show filamentary
small-scale structures across the entire sky. These features have narrow
line widths and were identified by \citet{Clark2014} and
\citet{Kalberla2016} as being aligned with similar structures in thermal dust
emission along the line of sight. By interpreting the interstellar medium
(ISM) as a multiphase medium with a cold neutral medium (CNM) in
pressure equilibrium with a diffuse warm neutral medium (WNM)
\citep[e.g.,][]{Wolfire2003}, such filaments have been identified as
signatures of the CNM. These structures are associated with cold dust
and stretched out along the magnetic field lines. The nature of these
dense cold structures and their association with far infrared (FIR) structures has been
further investigated by \citet{Clark2019}, \citet{Peek2019},
\citet{Clark2019b}, \citet{Murray2020}, \citet{Kalberla2020},
\citet{Kalberla2021}, \citet{Kalberla2023}, and
\citet{Lei2024}. Filaments have been found to be associated with particular
cold CNM, an increased CNM fraction, and an enhanced FIR emissivity
$I_{\mathrm{FIR}}/N_{\mathrm{HI}}$ and understood as coherent
\hi\ fibers with local density enhancements in position-velocity space.

The interpretation of such elongated structures that can have large
aspect ratios as caustics originating from real physical structures was
questioned by \citet{Lazarian2018} due to it being inconsistent with
\citet{Lazarian2000}. Their theory predicts such structures as being velocity
caustics caused by turbulent velocity fluctuations along the line of
sight. In this case, the observed emission originates from different
volume elements well separated along the line of sight. Based on this
velocity crowding concept, \citet{Yuen2021} developed the velocity
decomposition algorithm (VDA) to prove this point of view.

Caustics are defined as structures that correspond to singularities of
gradient maps \citep[e.g.,][]{Castrigiano2004}. Investigating the theory
of caustics and velocity caustics in detail, \citet{Kalberla2024} has
shown that the VDA velocity caustics concept, involving velocity
crowding, lacks consideration of such singularities and therefore does not
help explain the alignment of FIR/\hi\ filaments. The so-called
velocity caustics are not actually caustics. The VDA data products may
contain caustics in the same way as the original
position-position-velocity data cubes, but VDA does not serve as a good
mathematical concept to explain the alignment of FIR/\hi\ filaments in
comparison to a Hessian analysis of the original observations.

Concerning the physics of the ISM, it was shown in
\citet{Kalberla2024} that caustics originate unambiguously from
dense CNM structures, hence real observable entities. \hi\ absorption
line data that intersect with filaments have been used to probe the physics
of the filaments. This method of analysis was made possible by using the BIGHICAT
database by \citet{Naomi2023} that at the time of publication combined
all publicly available \hi\ absorption data. \citet{Kalberla2024} have shown that the CNM
in the diffuse ISM is exclusively located in filaments, caustics with
FIR counterparts.

BIGHICAT comprises 372 unique lines of sight distributed all over the
sky \citep[see][Fig. 3]{Naomi2023}, but only recently \citet{Nguyen2024}
provided absorption line data from the GASKAP-\hi\ survey on 2 714
positions. These measurements from the Pilot Phase II Magellanic Cloud
\hi\ foreground observations cover a 250-square-degree area of the Milky
Way foreground toward the Magellanic Clouds.  This region is
characterized by the intersection of two prominent filaments, composed
of gas and dust. Within the current sensitivity limitations, 462 lines of
sight, predominantly covering filamentary structures, show significant
absorption lines \citep[][Fig. 1]{Nguyen2024}.\footnote[1]{The 
  GASKAP-\hi\ main survey in full mode, with an integration time 20 times
  longer than the Pilot II observations considered here, will achieve a $\sim  4.5$ times higher
  sensitivity}

The focus of this paper is on these new additional absorption line
data, and we intend to extend the analysis in Sect. 7 of
\citet{Kalberla2024}. We use observational parameters from
\citet{Nguyen2024} to derive the physical nature of FIR/\hi\ caustics
that are associated with absorption components. Additional
  parameters (e.g., CNM Doppler temperature, volume density, and
  turbulent Mach number) are derived directly from the fitted CNM
  parameters, as discussed in more detail by Lynn et al. (2025, in
  preparation). In this work, we present in Sect. \ref{Obs} a brief
summary of observations and data reduction. In Sect. \ref{Results} we
report the derived results and discuss them in Sect. \ref{Discussion}.

\begin{figure*}[ht] 
   \centering
   \includegraphics[width=8.5cm]{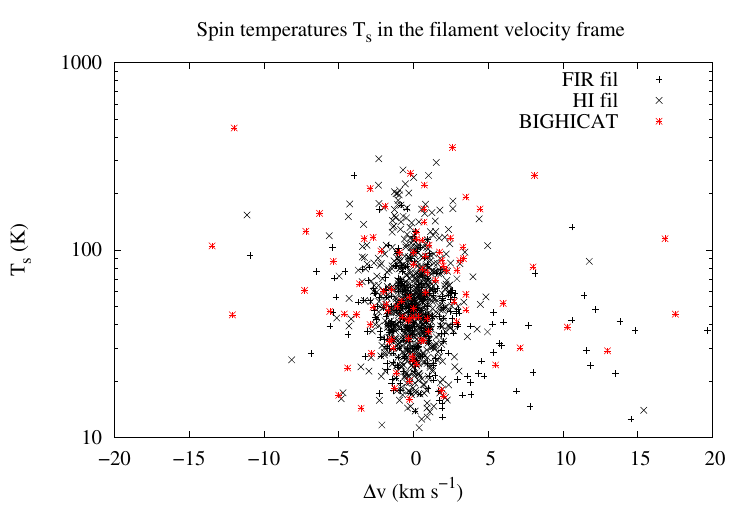}
   \includegraphics[width=8.5cm]{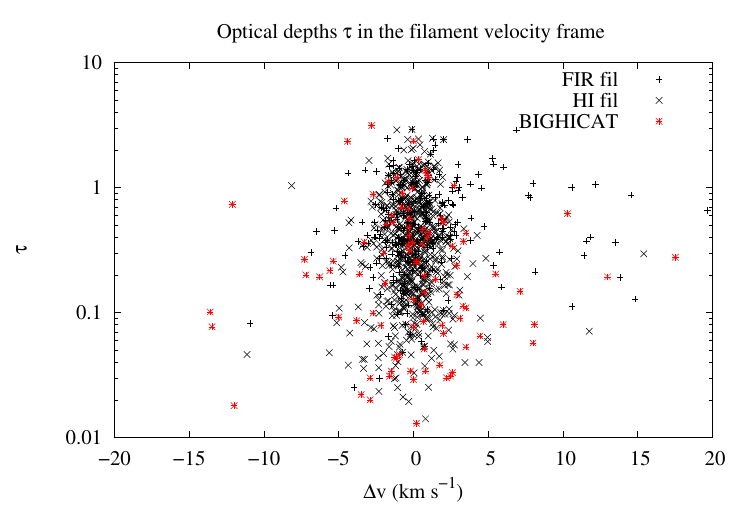}
   \includegraphics[width=8.5cm]{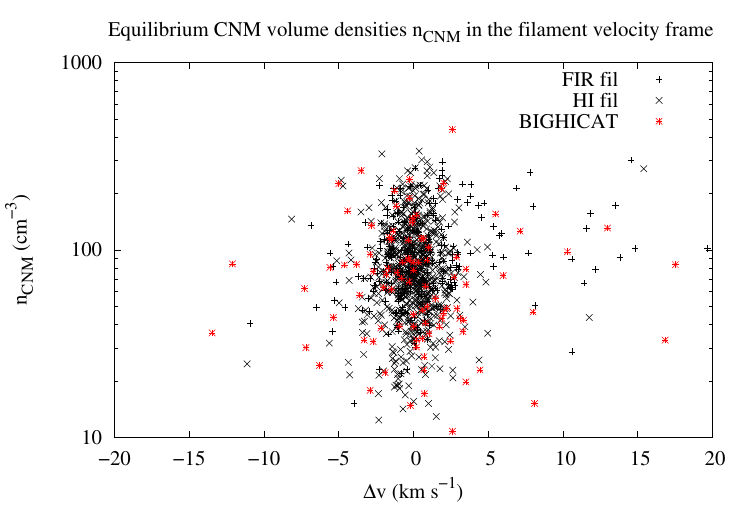}
   \includegraphics[width=8.5cm]{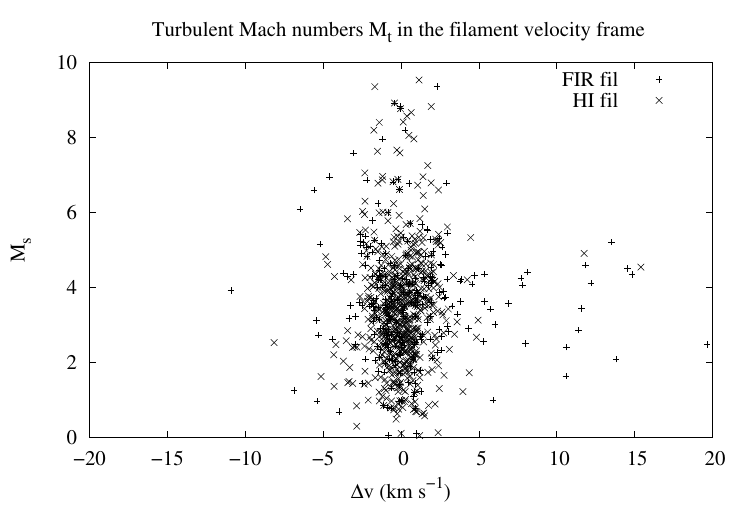}
   \includegraphics[width=8.5cm]{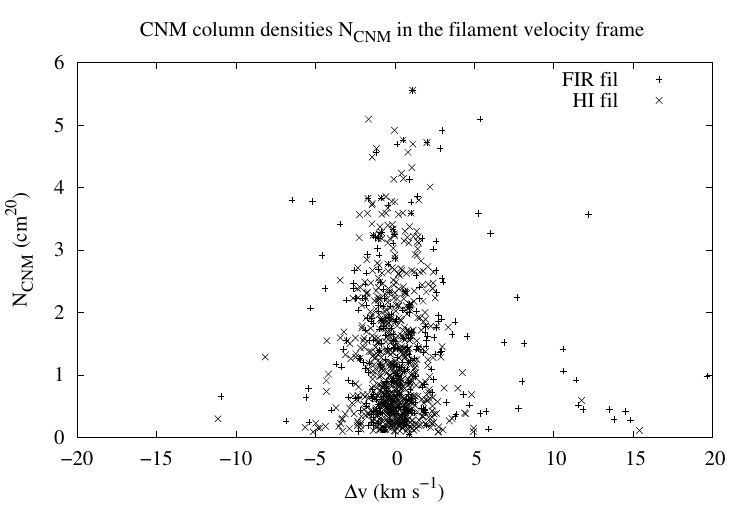}
   \includegraphics[width=8.5cm]{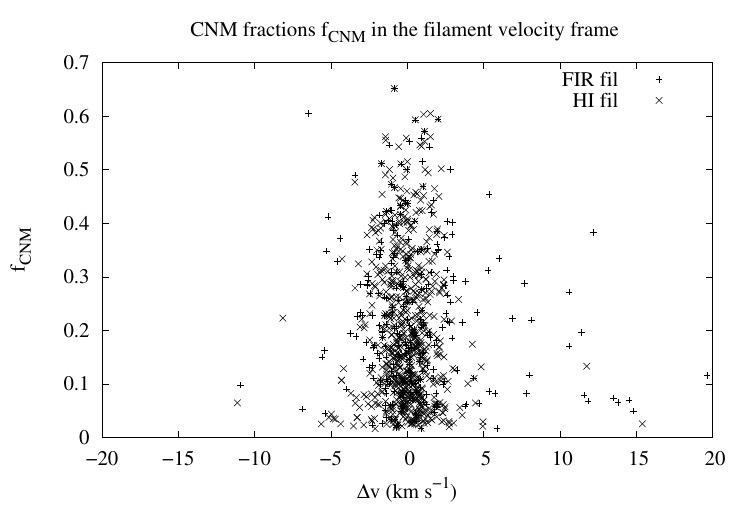}
   \caption{Derived physical parameters for absorption components
     ($T_\mathrm{s}$, $\tau$, $n_\mathrm{CNM}$, $M_\mathrm{t}$,
     $N_\mathrm{CNM}$, and $f_\mathrm{CNM}$ ) depending on turbulent
     velocity deviations $ \Delta v$ from filament velocities. Data from
     the GASKAP-\hi\ survey are indicated by black symbols; + in case of
     FIR/\hi\ filaments and x in case of genuine
     \hi\ filaments. BIGHICAT data from FIR/\hi\ filaments at latitudes
     $|b| > 10 \deg$ are included for reference and indicated by red
     symbols.  }
   \label{Fig_vfil}
\end{figure*}

\section{Observations and data reduction}
\label{Obs}

For the characterization of filaments, we considered these \hi\ structures
as caustics or local singularities in the data distribution. We used a
Hessian analysis \citet{Kalberla2021}, \citet{Kalberla2023}, and
\citet{Kalberla2024} (hereafter Paper I, II, and III) to calculate these
singularities. Caustics in smooth differentiable maps are related to
critical points, positions where the derivative of the investigated
distribution is zero \citepalias{Kalberla2024}. In 2D maps, caustics are
defined by maxima that can be identified as minima in the Hessian
eigenvalue distribution. The orientations of the caustics are given by
the corresponding eigenvectors or orientation angles, which are more
convenient for our applications. The Hessian analysis was applied in
\citetalias{Kalberla2021} and \citetalias{Kalberla2023} to {\it Planck}
FIR data at 857 GHz \citep{Planck2020}. We have used also HI4PI \hi\ observations
\citep{HI4PI2016}, which combine data from the Parkes Galactic all sky
survey  (GASS; \citealt{Naomi2009}, \citealt{Kalberla2015}) and the
  Effelsberg-Bonn \hi\ survey (EBHIS; \citealt{Winkel2016}).  Velocity channels in the range $ -50 < v_{\mathrm{LSR}} < 50 $
  \kms were considered.  For details, we refer to
\citetalias{Kalberla2021} and \citetalias{Kalberla2023}.

We used the publicly available results from the GASKAP-\hi\ survey by
\citet{Nguyen2024} to check whether the observed absorption components are
correlated with caustics (see Appendix \ref{Appendix}). For each observed
position, we first checked whether it is located along a known
  FIR/\hi\ filament \citepalias{Kalberla2021}. If this was the case, we
  determined the eigenvalue $\lambda_{-}^{\mathrm{FIR}}$ at this position
  \citepalias[][Eq. (2)]{Kalberla2021} and the velocity of the
  associated FIR caustic $v_\mathrm{fil}^{\mathrm{FIR}}$ and assigned it
  to the absorption component. In the case of multiple absorption components
  at the same position, we selected the component with the smallest
  velocity deviation $ |\Delta v| = |v_\mathrm{abs} -
  v_\mathrm{fil}^{\mathrm{FIR}}| $. In \citetalias{Kalberla2021} it was
  shown that FIR/\hi\ filaments are shaped by a small-scale turbulent
  dynamo. In such an environment, $\Delta v$ defines the turbulent
  velocity deviation of the CNM condensations (absorption features) from the
  main body (average) of the filament, that is observed in emission.
    
  For all the absorption components, we searched
  for a correlation between the absorption feature with an 
  \hi\ filament. The relation between FIR/\hi\ and associated
  \hi\ filaments in channel maps is discussed in detail in Sect. 7 of
  \citetalias{Kalberla2024}.  We determined the minimum eigenvalue
  $\lambda_{-}^{\mathrm{HI}}$ and the related filament (or critical
  point) velocity $v_\mathrm{fil}^{\mathrm{HI}}$ from the \hi\ database.
  In the case of multiple absorption components at the same position, we
 also used $ \Delta v = v_\mathrm{abs} -
  v_\mathrm{fil}^{\mathrm{HI}}$ for the closest match in velocity.

 Figure \ref{Fig_J032548} in the Appendix illustrates the
 relation between caustics and \hi\ absorption features in two cases.
 Altogether, we found that all the 462 GASKAP-\hi\ lines of sight with
 absorption intersect filaments. Of these, 262 positions are located at
 FIR caustics, and all 691 absorption components are associated
 with caustics in \hi.

\section{Results}
\label{Results}

The magnitude of the eigenvalue $-\lambda_{-}$ from the Hessian analysis
is a measure of the strength and significance of a filamentary
structure. We refer to Fig. \ref{Fig_lam} in the Appendix
  for details.  For turbulent velocity deviations $
\Delta v$, we determined in case of GASKAP-\hi\ data an average
  $v_\mathrm{av} = 6.~ 10^{-2}  $ \kms\ with a dispersion of $\delta
  v_\mathrm{turb} = 2.48 $ \kms. In the case of FIR filaments, we obtained a larger
  dispersion of $\delta v_\mathrm{turb} = 3.9 $ \kms. This dispersion
along the line of sight is consistent with a mean velocity dispersion of
$\delta v_\mathrm{turb} = 3.8 $ \kms\ along the
FIR/\hi\ filaments projected to the plane of the sky, hence
perpendicular to the line of sight \citepalias{Kalberla2024}.
Internal velocity dispersions for individual filaments tend to depend on
the aspect ratio of the filaments. For the prominent filament that is
crossing the GASKAP-\hi\ field of view, $\delta v_\mathrm{turb} \sim 5.7
$ \kms\ was estimated previously. The GASKAP-\hi\ positions, however, only cover
a small fraction of this interesting filament that was
shown in Fig. 17 of \citetalias{Kalberla2023}.

The primary physical parameters that characterize the thermodynamic
properties of the CNM are spin temperature $T_\mathrm{s}$ and volume
density $n_\mathrm{CNM}$.  The most critical parameters from
observations are the optical depth $\tau$ and column density
$N_\mathrm{CNM}$, as they determine the significance of derived results. The
dynamics of the CNM are characterized by the turbulent Mach number
$M_\mathrm{t}$, and the cold gas fraction $f_\mathrm{CNM}$ characterizes
the composition of the multiphase medium. We used the parameters
$T_\mathrm{s}$, $\tau$, $N_\mathrm{CNM}$, and $f_\mathrm{CNM}$ as
determined and tabulated by \citet{Nguyen2024} but additionally determined
the volume densities $n_\mathrm{CNM}$ and Mach numbers $M_\mathrm{t}$ 
  from the fitted CNM Gaussian components. The distribution of these
parameters as a function of turbulent velocity deviations $ \Delta v$ are
shown in Fig. \ref{Fig_vfil}. For $T_\mathrm{s}$, $\tau$, and
$n_\mathrm{CNM}$, we added BIGHICAT data for comparison.

For the determination of volume densities $n_\mathrm{CNM}$, we assumed
that the CNM is in thermal equilibrium with the associated WNM at a
characteristic equilibrium pressure of $\log (p/{\mathrm{k}}) = 3.58$
\citep{Jenkins2011}. This applies if the CNM filaments had sufficient
time to reach thermal equilibrium, a condition that most probably is
valid in magnetically aligned filaments
\citepalias[][Sect. 7.1]{Kalberla2024}. As $n_\mathrm{CNM} \propto
T_\mathrm{s}^{-1} $, the volume density distribution in
Fig. \ref{Fig_vfil} is inverse to the $ T_\mathrm{s} $ distribution. For
a median $ T_\mathrm{s} = 50 $ K, we obtained a median density of
$n_\mathrm{CNM} = 76 ~ \rm cm^{-3}$.

\begin{figure}[ht] 
   \centering
   \includegraphics[width=8.5cm]{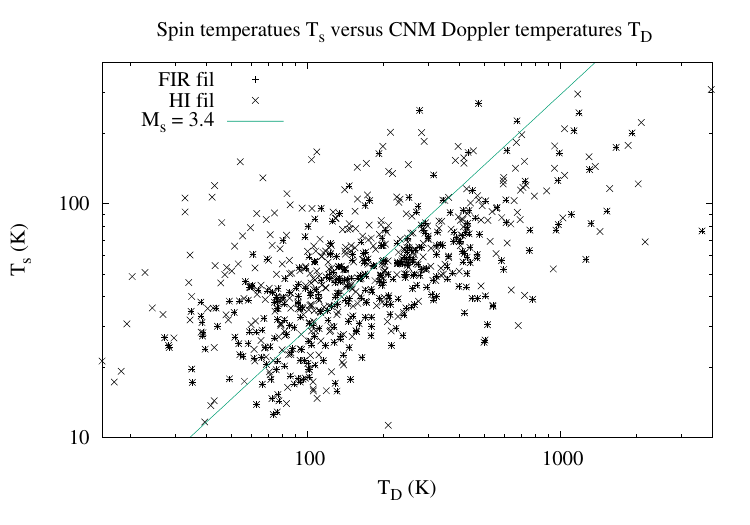}
   \caption{Spin temperatures $T_\mathrm{s}$ versus Doppler temperatures
     $T_\mathrm{D}$ derived from GASKAP-\hi\ data. A constant slope for
     $ M_\mathrm{t} = 3.4 $ is indicated. }
   \label{Fig_TD}
\end{figure}

To determine the turbulent Mach numbers, we used the relation
$M_\mathrm{t} = \sqrt{(4.2(T_\mathrm{D}/T_\mathrm{s} -1)}$
\citep[e.g.,][]{Heiles2003}.  Here, $T_\mathrm{D}$ is the Doppler temperature,
determined as $T_\mathrm{D} = 121 ~ \sigma^2_\mathrm{em} $ from the
velocity dispersion $\sigma_\mathrm{em} $ of the observed
GASKAP-\hi\  CNM component in emission. In a turbulent ISM, we have an upper
limit to the kinetic temperature $T_\mathrm{D} > T_\mathrm{k}
$,\footnote[2]{A few authors therefore denote $T_\mathrm{D}$ as
  $T_{\mathrm{k,max}}$.} and for the CNM, we can use the approximation
$T_\mathrm{D} > T_\mathrm{s} $ \citep{Naomi2023}. In the absence of optical
depth measurements, $T_\mathrm{D}$ is often used to estimate
temperatures of the \hi\ gas. From the GASKAP-\hi\ database, we determined
an average $ M_\mathrm{t} = 3.4 \pm 1.6 $ for all components but $
  M_\mathrm{t} = 3.7 \pm 1.6 $ for those associated with FIR
  filaments; the latter is consistent with the results from the
  Millennium Survey by \citet{Heiles2005}.  Figure \ref{Fig_TD} shows the
relation between $ T_\mathrm{D} $ and $ T_\mathrm{s} $. The huge scatter
is obvious, and $ T_\mathrm{D} $ does not measure a temperature, but
it is still safe to use $ T_\mathrm{D} $ to derive the nature of a
multiphase component. For example, it is valid to assume that filaments with
typical Doppler temperatures $ \textlangle T_\mathrm{D} \textrangle \sim
220 $ K are dominated by the CNM \citep{Clark2019} or
\citep{Kalberla2016}. The GASKAP-\hi\ data, represented in
Fig. \ref{Fig_TD}, are important since they provide, for the first time, a
large homogeneous database that is useful to observationally relate $
T_\mathrm{D} $ with $ T_\mathrm{s}$.

The source distribution, observed by \citet{Nguyen2024}, enables the small-scale structure of the CNM to be studied in position-velocity
space. For each of the observed absorption positions, we determined the
nearest neighbors. Then, we compared pairwise absorption components
  with the least velocity deviations $|\Delta v|$. Similar to
\citet{Crovisier1985}, we determined relative parameter changes defined as
\begin{equation}
     \label{eq:diff} 
\Delta P (x) = \frac{| P(x_1) - P(x_2) | } {P(x_1) + P(x_2) }.
\end{equation}  
Here, P can be replaced with $ T_\mathrm{s} $, $\tau$, or any of the measured
absorption line parameters depending either on separation $S$ in position
or displacement in velocity $\Delta v = v_1 - v_2 $.

\begin{figure}[ht] 
   \centering
   \includegraphics[width=8.5cm]{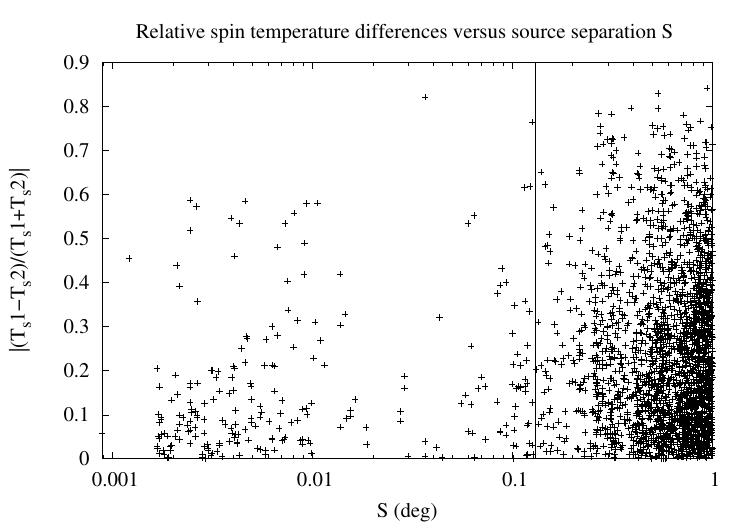}
   \includegraphics[width=8.5cm]{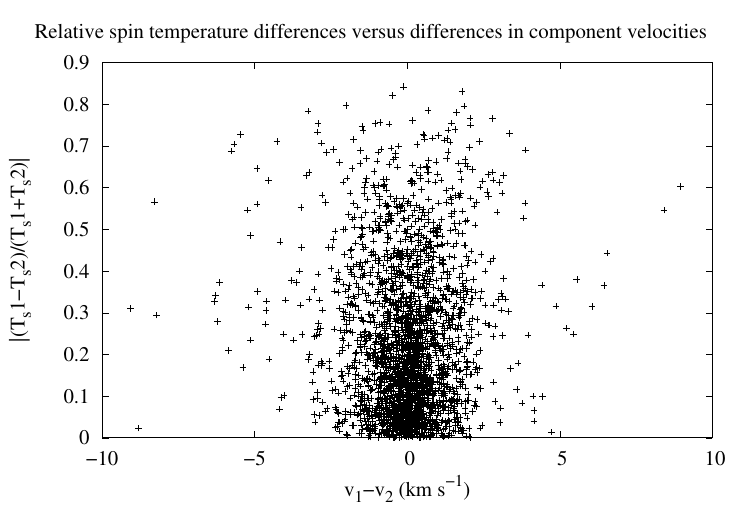}
   \caption{Relative changes $\Delta T_\mathrm{s}$ in spin temperatures
     $T_\mathrm{s}$. Top: Relative changes $\Delta T_\mathrm{s}$ depending on source 
       separations $S$. The vertical line indicates the average
     filament width. Bottom: Relative changes $\Delta T_\mathrm{s}$ depending on velocity
     differences $ v_1 - v_2 $. }
   \label{Fig_diffdist}
\end{figure}

Figure \ref{Fig_diffdist} shows, as an example, the relative changes in spin
temperature $ T_\mathrm{s} $ depending on the displacement and velocity
spread of the component pairs.\footnote[3]{In all cases, the pairs are
    defined as the absorption components with the minimum velocity
    deviation $|v_1 - v_2 |$} For angular separations of up to 1\deg, we
  determined the relative parameter changes $\Delta T_\mathrm{s} = 0.23 \pm
  0.18$, $\Delta \tau = 0.31 \pm 0.22 $, $\Delta N_\mathrm{CNM} = 0.28
  \pm 0.21$, and $\Delta f_\mathrm{CNM} = 0.26 \pm 0.20 $. The scatter is large for each of
  the parameters, but on average the relative
  changes are moderate.

\begin{figure}[ht] 
   \centering
   \includegraphics[width=8.5cm]{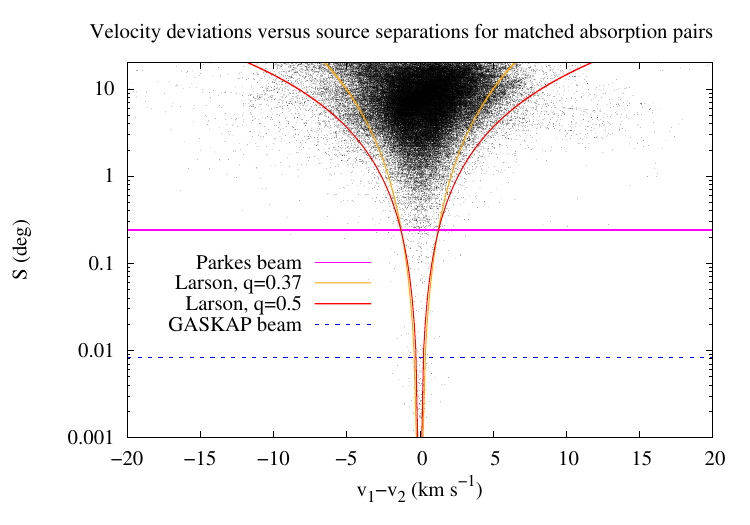}
   \caption{Velocity differences $ v_1 - v_2 $ depending on angular
     source separation $S$ for matched absorption components. The
     horizontal lines indicate the Parkes and GASKAP beamwidths.  The
     red and yellow lines display envelopes to the distribution,
     assuming a turbulent scaling between velocity and length scale
     \citep{Larson1979} with exponent $q$ (see text).  }
   \label{Fig_Larson}
\end{figure}

To gain insight into the nature of the observed velocity fluctuations,
we considered the deviations $\Delta v $ for all pairs of matching absorption
components for neighbor positions of up to
20\deg\ of separation. Fig. \ref{Fig_Larson} shows a clear trend that
$\Delta v $ increases systematically with separation. A key feature of
turbulence is that motions are spatially correlated
\citep{Larson1979}. Following \citet[][Eq. (1)]{Wolfire2003}, we used the
relation $\sigma_\mathrm{v} (l) = \sigma_\mathrm{v}(1)\, l^q$, where $l$ is
the length scale and $\sigma_\mathrm{v} (1) = 1.2 $ \kms\ is the
dispersion at $l= 1$ pc. In the case of supersonic turbulence, $q$ is
expected to be in the range $1/3 \la q \la 1/2$; \citet{Larson1979}
found $q = 0.37$. In Fig. \ref{Fig_Larson}, we plot for comparison the
solutions for $q = 0.37$ and $q = 0.5$, assuming that the CNM is at a
distance of 250 pc \citep{Nguyen2024}.  Consistent with
\citet{Heyer2009}, $q \sim 0.5$ appears to match the data best over a
length scale of four orders of magnitude, with only an insignificant
number of outliers.

\section{Discussion}
\label{Discussion}

Sensitive \hi\ absorption data are essential for understanding the
physics of the ISM. The recent investigations by \citet{Nguyen2024}, which represent the
largest Galactic neutral hydrogen \hi\ absorption survey to date, more
than double the available absorption lines at high Galactic latitudes.
We used the physical parameters derived by these authors and related them
to caustics, filamentary structures that have been determined previously
from emission line observations in \hi\ and in FIR observations by {\it
  Planck}.

The absorption line detections by \citet{Nguyen2024} probe the CNM, and
these lines of sights intersect FIR and \hi\ filaments. Of
the positions, 57\% are found to be related to common FIR/\hi\ structures with
coherent filament velocities. To compare \hi\ and FIR structures, we
  needed to take sensitivity limitations into account. It was shown in
Figs. 8 and 9 of \citetalias{Kalberla2024} that only a fraction of the
\hi\ filaments can be assigned to FIR structures, but still there is a
close relation between FIR and \hi.  The center velocities of the
observed absorption lines share the filament velocities, observed in
\hi\ emission, with some fluctuations. These velocity deviations, which
we address to turbulent motions, have a dispersion of $\delta
  v_\mathrm{turb}^{\mathrm{HI}} = 2.48 $ \kms\ for the whole
  \hi\ sample, but $\delta v_\mathrm{turb}^{\mathrm{FIR}} = 3.9 $
  \kms\ for FIR filaments. The FIR filaments tend to be more prominent
  with larger aspect ratios; hence, they show a larger dispersion
  \citepalias{Kalberla2024}.  The FIR/\hi\ filaments tend to have lower mean
  spin temperatures, $T_\mathrm{s}^\mathrm{FIR} \sim 49 $ K, compared to
  components that are only associated with \hi\ filaments,
  $T_\mathrm{s}^\mathrm{HI} \sim 67 $ K. Thus, dusty FIR filaments appear
  to be colder but more turbulent. All of the GASKAP-\hi\ absorption
  components are located within caustics. This finding supports our
previous conclusions from BIGHICAT data that the FIR/\hi\ filaments
originate from dense CNM structures \citepalias{Kalberla2024}.

Cold neutral medium components that are observed in absorption trace the filaments
in position-velocity space. For an estimated distance of 250 pc
\citep{Nguyen2024}, the typical FIR/\hi\ filament width is 0.63 pc. A
GASKAP beam corresponds to a linear size of $\sim 0.04 $ pc.
For an observed dispersion of $\Delta v \sim 4 $ \kms, the filament crossing
time for small-scale CNM structures is typically 0.16 My, which is large in
comparison to the expected cooling time of $ 3 \cdot 10^4$ y
\citep{Jenkins2011}. In the case of phase transitions for \hi\ gas with low
turbulent deviation velocities, condensations with a high CNM fraction
$f_\mathrm{CNM}$ (Fig. \ref{Fig_vfil}) have time to develop.

\begin{figure}[ht] 
   \centering
   \includegraphics[width=8.5cm]{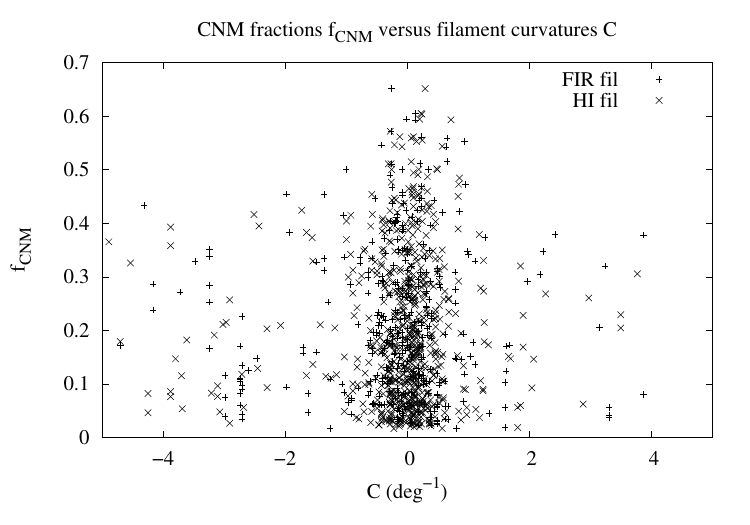}
   \caption{Cold neutral medium fractions $f_\mathrm{CNM}$ depending on local filament
     curvatures $C$. }
   \label{Fig_curv}
\end{figure}

We considered the question of whether any of the derived CNM parameters, in
particular spin temperature $T_\mathrm{s}$, optical depth $\tau$, and CNM
fraction $f_\mathrm{CNM}$, could depend on the apparent strength of the
filamentary structure, as characterized by its eigenvalue
$-\lambda_{-}$. We find no obvious dependencies (see
  Fig. \ref{Fig_J032548} as an example for a deep critical point that is
  associated with a low optical depth feature). In particular, we found
  that caustics at a level of $\lambda_{-} \la -5~ \mathrm{K deg}^{-2}$, which is ten times lower than assumed in  \citetalias{Kalberla2021}, are still
  significant.

  Considering the fluctuations of the derived physical parameters, we found
  on average only moderate relative changes of 23\% to 31\% in
  $T_\mathrm{s}$, $\tau$, or  other parameters, independent of source
separation or velocity deviation. However, we found indications
that the observed dense CNM structures with significant optical depth
exist preferentially at filament locations with a low local filament
curvature $C = 1/R $ and a large local radius $R$ (see Fig. \ref{Fig_curv}
for $f_\mathrm{CNM}$ as an example).

Filaments with curvatures of $|C| \la 1 ~\mathrm{deg}^{-1} $ can, as
demonstrated by \citet{Clark2014}, be straightforwardly analyzed with
the Rolling Hough Transform with window diameters $50 \arcmin \la D_\mathrm{W}
\la 125\arcmin$. For filaments that are aligned with the magnetic field
line, a low curvature is expected at positions with an enhanced magnetic
field. For a small-scale dynamo, magnetic field strength and curvature are anticorrelated, $|B| \propto |C|^{-1/2} $
(\citealt{Schekochihin2004} and \citealt{St-Onge2018}), and the curvature
distribution of FIR/\hi\ filaments was found to be characteristic for
structures generated by a small-scale dynamo \citepalias{Kalberla2021}.
As demonstrated with Fig. \ref{Fig_TD}, these \hi\ filaments that have previously been observed in emission with
characteristic Doppler temperatures of $T_\mathrm{D} = 220 $ K belong to the CNM.

In summary, we conclude that all currently available data from
\hi\ absorption line observations at high Galactic latitudes are
consistent with the assumption that a major part, if not all, of the CNM
is located within magnetically aligned
filaments that can be traced on large scales in FIR and in
\hi\ emission. These FIR/\hi\ caustics are affected by turbulent
velocity perturbations of 3 to 4 \kms. The filaments are coherent structures
in position-velocity space with well-defined physical properties as CNM
in a diffuse multiphase medium.


\begin{acknowledgements}
This scientific work uses data obtained from Inyarrimanha Ilgari Bundara
/ the Murchison Radio-astronomy Observatory. We acknowledge the Wajarri
Yamaji People as the Traditional Owners and native title holders of the
Observatory site. CSIRO’s ASKAP radio telescope is part of the Australia
Telescope National Facility (https://ror.org/05qajvd42). Operation of
ASKAP is funded by the Australian Government with support from the
National Collaborative Research Infrastructure Strategy. ASKAP uses the
resources of the Pawsey Supercomputing Research Centre. Establishment of
ASKAP, Inyarrimanha Ilgari Bundara, the CSIRO Murchison Radio-astronomy
Observatory and the Pawsey Supercomputing Research Centre are
initiatives of the Australian Government, with support from the
Government of Western Australia and the Science and Industry Endowment
Fund.  HI4PI is based on observations with the 100-m telescope of the
MPIfR (Max-Planck- Institut für Radioastronomie) at Effelsberg and
Murriyang, the Parkes radio telescope, which is part of the Australia
Telescope National Facility (https://ror.org/05qajvd42) which is funded
by the Australian Government for operation as a National Facility
managed by CSIRO. This research has made use of NASA's
Astrophysics Data System.  Some of the results in this paper have been
derived using the HEALPix package.
   \end{acknowledgements}

\begin{appendix}

\section{Eigenvalues}
\label{Appendix}

Eigenvalues $\lambda_{-}$ are determined from a Hessian analysis of the
HI4PI survey according to \citetalias{Kalberla2024}, Eqs. (1) and
(2). Caustics can be identified as critical points with $\lambda_{-} <
0$. These eigenvalues have been determined all sky for velocities $ -50
< v_{\mathrm{LSR}} < 50 $ \kms. To distinguish \hi\ and FIR caustics,
orientation angles $\Theta$ according to \citetalias[][Eq. (3)]{Kalberla2024},
were used.  At a given position, the line of sight may
intersect several \hi\ caustics at different velocities but only a
single FIR caustic can be detectable. In addition to eigenvalues, the
associated orientation angles (derived from eigenvectors) are used to
match FIR and \hi\ filaments.  The \hi\ caustic with the best
alignment, hence the minimum of $| \Theta^\mathrm{FIR} -
\Theta^\mathrm{HI} |$, defines the velocity of the FIR filament. 
The uncertainties in matching orientation angles of FIR and \hi\ filaments
are below 3\degr, see \citetalias[][Table 1]{Kalberla2021}.

\subsection{Eigenvalue spectra}
\label{App1}

\begin{figure}[th] 
   \centering
   \includegraphics[width=9cm]{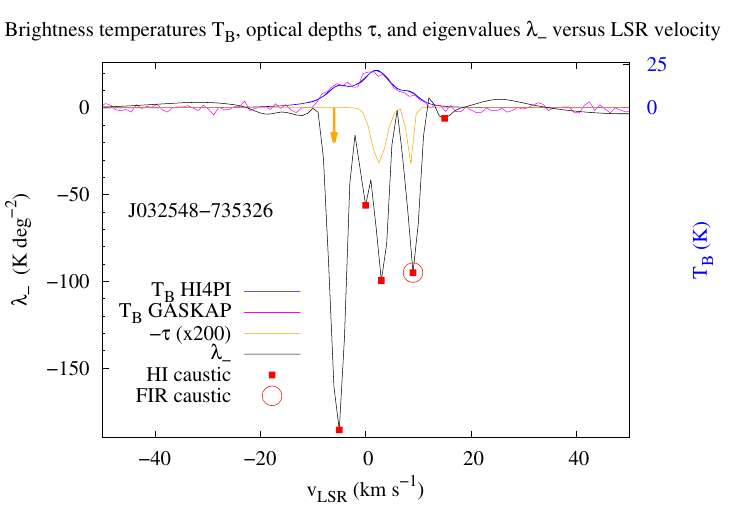}
   \includegraphics[width=9cm]{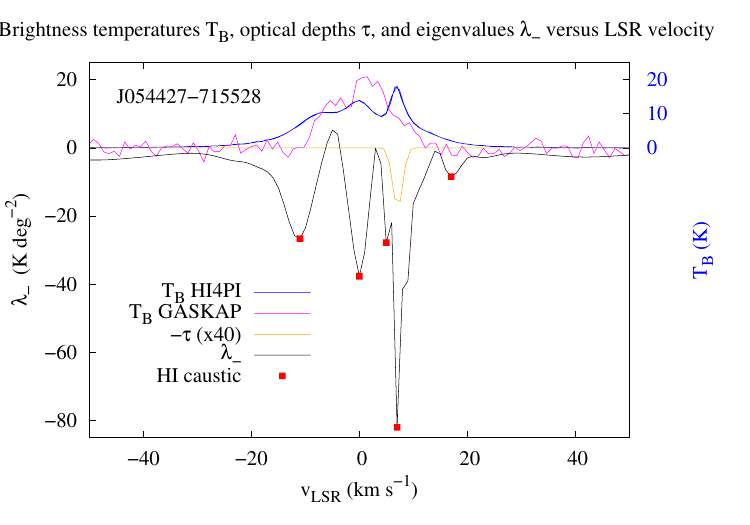}
   \caption{Hessian eigenvalues $\lambda_{-}$ (black) related to
     observed brightness temperatures at the nearest position. HI4PI
     data are indicated in blue and GASKAP-\hi\ data in magenta.
     Critical points $\lambda_{-} < 0$ are marked in red. Top: for the
     source J032548-735326 two absorption components (orange, scaled by
     a factor of 200) are indicated with an additional feature below the
     detection limit (arrow).  For observational details see Fig. 8 of
     \citet{Nguyen2024}.  Bottom: the source J054427-715528 has a single
     absorption component (orange, scaled by a factor of 40). For
     observational details see Fig. 3 of \citet{Nguyen2024}. }
   \label{Fig_J032548}
\end{figure}

 Figure \ref{Fig_J032548} shows two examples for the relation between
 eigenvalues $\lambda_{-}$ and \hi\ absorption.  On top of
 Fig. \ref{Fig_J032548} the source J032548-735326 is shown. The
 brightness temperatures, observed with HI4PI and GASKAP-\hi\ close to
 the source, are plotted in blue and magenta. The eigenvalues
 $\lambda_{-}$ (black) indicate five \hi\ caustics (red), one of them is
 associated with an FIR caustic. From the GASKAP-\hi\ pilot project
 observations, two significant \hi\ absorption lines (yellow) can be
 identified; a third potential absorption feature (arrow) is observed
 below the detection threshold. These absorption features are associated
 with caustics, for more observational details see
 \citet[][Fig. 8]{Nguyen2024}.

The source J054427-715528, shown at the bottom of
Fig. \ref{Fig_J032548}, has five \hi\ caustics that are not related to
an FIR filament. Only one of these caustics is associated with a
detected absorption line. However, we observed significant fluctuations at
velocities of two of the caustics when comparing the HI4PI and
GASKAP-\hi\ brightness temperature profiles close to the source
position. The HI4PI and GASKAP-\hi\ beamwidths (14\farcm5 and 30\arcsec,
respectively) are largely different, hence these fluctuations indicate
small scale structure that is usually addressed to CNM. 
More observational details for the source J054427-715528 are given in
Fig. 3 of \citet{Nguyen2024}.

The example eigenvalue spectra in Fig. \ref{Fig_J032548} have been
chosen to allow a comparison with published absorption spectra. Depending
on the \hi\ emission distribution, the $\lambda_{-}$ spectra can have
far more complex shapes. For the data analysis it is only necessary to
determine critical points with $\lambda_{-} < 0$, marked in red in the
examples\footnote[4]{For downloads of eigenvalue spectra see \url{https://www.astro.uni-bonn.de/hisurvey/AllSky_gauss/index.php}}.

\subsection{Turbulent velocity dispersions}
\label{App2}

In \citetalias{Kalberla2021} it was shown that FIR/\hi\ filaments are
shaped by a small-scale turbulent dynamo. In such a case the filament
curvatures show a characteristic pattern (\citealt{Schekochihin2004} and
\citealt{St-Onge2018}). A succession of random velocity shears stretches
the magnetic field and leads on the average to its growth to dynamical
strengths. This process leads to filamentary structures of near-constant
velocity along magnetic lines of force. Both, characteristic curvatures
and near-constant velocities along filaments, are observed
\citepalias{Kalberla2021}.  Numerical studies of the condensation of the
WNM into CNM structures under the effect of turbulence and thermal
instability by \citet{Saury2014} indicate that the velocity field of the
CNM reflects the velocity dispersion of the WNM. These authors estimate
a CNM cloud-to-cloud velocity dispersion to 5.9 \kms\ from observations
and find about 3 \kms\ in simulations.

In case of phase transitions within a turbulent medium that is affected
by a dynamo it is expected that the near-constant velocity field along
magnetic lines of force is locally perturbed. Assuming the CNM aligned
preferential along the filaments, a CNM cloud-to-cloud velocity
dispersion in the order of 3 to 6 \kms\ is expected along the filaments
\citepalias[][Sect. 7]{Kalberla2024}.  In this context it is also
reasonable to assume turbulence along the line of sight and to define a
turbulent velocity deviation $\Delta v$ between filament and absorption
component.  The GASKAP and GASS beams are very different, thus $\Delta
v$ measure differences between pencil beam of 30\arcsec and the
effective beam of the Hessian operator (5x5 pixels, or
$\sim$18\arcmin\ for an nside=1024 HEALPix grid) which is close to the
GASS beam. Thus, $\Delta v$ measures in this case the velocity
deviations of absorption components relative to the filament body.

\begin{figure}[ht] 
   \centering
   \includegraphics[width=8.5cm]{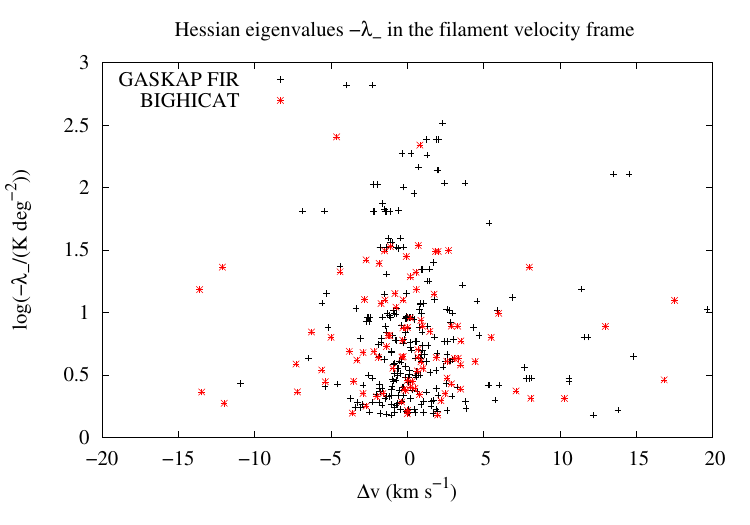}
   \includegraphics[width=8.5cm]{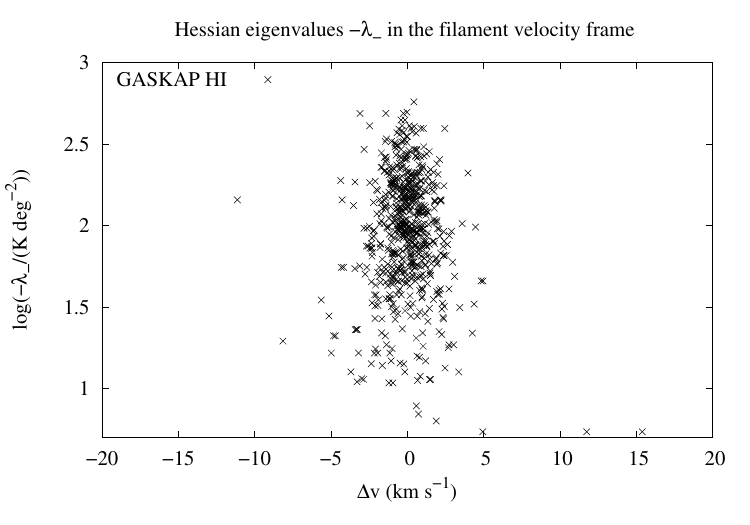}
   \caption{Hessian eigenvalues $\lambda_{-}$ for absorption components
     depending on turbulent velocity deviations $ \Delta v$ from
     filament velocities. Top: GASKAP-\hi\ data for FIR filaments,
     including BIGHICAT results. Bottom: GASKAP-\hi\ data for
     \hi\ filaments, no BIGHICAT data in this case. }
   \label{Fig_lam}
\end{figure}

In Fig. \ref{Fig_lam} we display dependencies between caustic
eigenvalues $-\lambda_{-}$ and turbulent velocity deviations $ \Delta v
= v_\mathrm{abs} - v_\mathrm{fil}^{\mathrm{HI}}$ and $ \Delta v =
v_\mathrm{abs} - v_\mathrm{fil}^{\mathrm{FIR}}$ for \hi\ and FIR
filaments respectively. Comparing \hi\ and FIR, we need to take
different survey sensitivities into account
\citep[e.g.,][Fig. 4]{Kalberla2016}; \hi\ and FIR eigenvalues need to be
treated separately. Figure \ref{Fig_lam} shows the
GASKAP-\hi\ eigenvalues as function of the associated turbulent velocity
deviations $ \Delta v$ for FIR and \hi\ surveys independently.

The upper part of Fig. \ref{Fig_lam} considers FIR eigenvalues
$-\lambda_{-}$ depending on turbulent velocity deviations $ \Delta
v$. For comparison with Fig. 5 of \citetalias{Kalberla2024} we include
previous results from BIGHICAT, 97 components for galactic latitudes
$|b| > 10 \degr$. For GASKAP FIR we determine a dispersion of $\delta
v_\mathrm{turb} = 3.9 $ \kms, consistent with $\delta v_\mathrm{turb} =
3.7 $ \kms\ in case of BIGHICAT sources (red). This compares well to
$\delta v_\mathrm{turb} = 3.8 \pm 0.1$ \kms\ for the velocity dispersion along
the FIR/\hi\ filaments observed in the plane of the sky 
\citepalias{Kalberla2024}.

For turbulent velocity deviations $ \Delta v$ in case of \hi\ filaments
(lower part of Fig. \ref{Fig_lam}) we determine an average
$v_\mathrm{av} = 6.~ 10^{-2} $ \kms\ with a dispersion of $\delta
v_\mathrm{turb} = 2.48 $ \kms.  This result confirms the visual
impression that \hi\ filaments have significant lower turbulent
velocity deviations than FIR dominated structures.
 
\end{appendix}
\end{document}